\documentclass{article}
\usepackage[T1]{fontenc} 
\usepackage[utf8]{inputenc} 
\usepackage{ismir,amsmath,cite,url}
\usepackage{graphicx}
\usepackage{color}
\usepackage{multirow}

\usepackage{lineno}
\nolinenumbers

\title{A dataset and Baselines for Measuring and Predicting the Music Piece Memorability}

\multauthor
{Li-Yang Tseng \hspace{1cm} Tzu-Ling Lin \hspace{1cm} Hong-Han Shuai} { \bfseries{Jen-Wei Huang \hspace{1cm} Wen-Whei Chang}\\
 National Yang Ming Chiao Tung University, Hsinchu, Taiwan\\
{\tt\small \{liyangtseng.ee10, tzulinglin.11 ,hhshuai, admsd.ee10, wwchang\}@nycu.edu.tw}
}

\def\authorname{L.-Y. Tseng, T.-L. Lin, H.-H. Shuai, J.-W. Huang, and W.-W. Chang}

\begin{document}

\maketitle
\begin{abstract}
Nowadays, humans are constantly exposed to music, whether through voluntary streaming services or incidental encounters during commercial breaks. Despite the abundance of music, certain pieces remain more memorable and often gain greater popularity. Inspired by this phenomenon, we focus on measuring and predicting music memorability. To achieve this, we collect a new music piece dataset with reliable memorability labels using a novel interactive experimental procedure. We then train baselines to predict and analyze music memorability, leveraging both interpretable features and audio mel-spectrograms as inputs. To the best of our knowledge, we are the first to explore music memorability using data-driven deep learning-based methods. Through a series of experiments and ablation studies, we demonstrate that while there is room for improvement, predicting music memorability with limited data is possible. Certain intrinsic elements, such as higher valence, arousal, and faster tempo, contribute to memorable music. As prediction techniques continue to evolve, real-life applications like music recommendation systems and music style transfer will undoubtedly benefit from this new area of research.
\end{abstract}

\section{Introduction}\label{sec:introduction}
Music memorability is essential and has a wide range of commercial applications. For instance, content creators and marketing teams can use unique visual aids or audio components to captivate target audiences and distinguish themselves from other information sources~\cite{alexomanolaki2007music, hecker1984music}. Sound logos, such as Netflix's iconic ``ta-dum,'' are designed to engage listeners and promote brand recognition. In the realm of cognition literature, numerous studies have sought to understand the factors that contribute to music memorability~\cite{snyder2000music,ramsay2018intrinsic,halpern2008effects,mullensiefen2012role}. For instance, \cite{halpern2008effects,mullensiefen2012role} bridged the gap between cognitive science and MIR by examining whether implicit or explicit memory for a single tune is impacted by the type of encoding task and variations in timbre, tempo and structure.

However, music memorability remains a relatively unexplored area, particularly from a data-driven standpoint. Research related to music memorability includes the study of involuntary musical imagery (INMI)~\cite{mccullough2015catching,jakubowski2017dissecting}, also known as ``earworms,'' which refers to the phenomenon where fragments of music become mentally lodged on repeat. For instance, Jakubowski et al. proposed a model that can determine whether a piece of music may induce the INMI effect by using statistical analysis and a random forest model \cite{jakubowski2017dissecting}. However, the mechanism of INMI differs from music memorability since the former is a passive process while the latter can be active, e.g., everyone remembers how to sing ``Happy Birthday,'' but the song may not qualify as an earworm. Another line of prior studies~\cite{isola2013makes, Khosla_2015_ICCV, dubey2015makes, shekhar2017show} investigating the intrinsic memorability of multimedia content have predominantly focused on computer vision, with their findings suggesting that data-driven approaches can effectively determine memorability levels. Motivated by these studies, we break new ground in exploring music memorability from a data-driven perspective by compiling a novel dataset and employing machine learning techniques.

\begin{figure*}
 \centerline{
 \includegraphics[width=2\columnwidth]{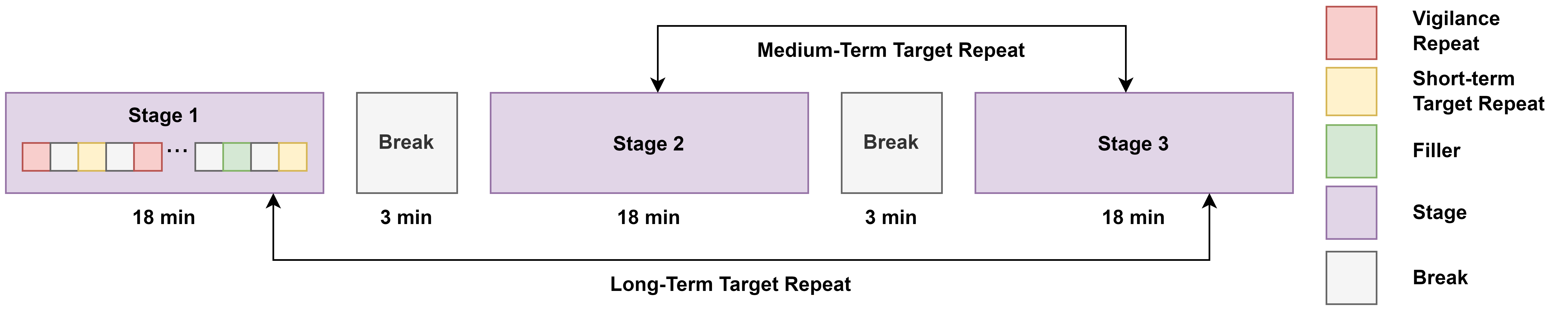}}
 \caption{The music memory game, which allows data annotators to label music memorability scores reliably. The experiment is divided into three stages, each with a 3-minute long break in between. Each 18-minute stage is composed of multiple 5-second music pieces and short breaks.}
 \label{fig:exp_overview}
\end{figure*}

Specifically, to expand the scope of memorability detection and recognition in music information retrieval (MIR), we establish a new research domain called music memorability regression (MMR), which aims to predict a memorability score for a given music piece. We create an experimental procedure as shown in \figref{fig:exp_overview} to collect a new dataset, the YouTube Music Memorability (YTMM) dataset, where memorability scores are determined by the percentage of participants who can recall the music piece after a certain period. This dataset provides reliable and consistent music memorability scores across all participants, paving the way for further research in the field. We also propose several baseline approaches for predicting music memorability, including feature engineering using hand-crafted music-related features and transfer learning techniques. These baselines not only demonstrate the potential of machine learning in addressing music memorability but also serve as a foundation for future work.

Despite the promise of machine learning in tackling music memorability by predicting memorability scores, its ``black box'' characteristics hinder the interpretation of machine decisions in MIR tasks. A straightforward approach would be to compute correlations without relying on black-box prediction models to glean insights about the relationship between memorability and musical features. However, given the complexity of analyzing music memorability, using a single feature results in an extremely low correlation with memorability, leading to inconclusive findings. One alternative would be to explore all possible feature combinations when calculating correlations, but the sheer number of combinations, e.g., $2^{20}-1$ for just 20 features, renders this approach impractical. A/B testing could be used to determine which type of music is more memorable, but it suffers from similar drawbacks, such as being time-consuming and unable to account for all variables that may impact the experiment's outcome. To make machine learning models reveal their ``black box'' characteristics, researchers are increasingly adopting explainable artificial intelligence (XAI)~\cite{adadi2018peeking} for deeper insights. Building on previous interpretability analyses in audio processing~\cite{li2020does, jeyakumar2020can}, we utilize Shapley Additive Explanations (SHAP)~\cite{NIPS2017_7062}, a game-theoretic approach that clarifies the output of machine learning models, to identify the key components of memorable music.

Our main contributions are as follows: first, we present the new YTMM dataset with objective annotations of memorability scores, which will be publicly available for future research; second, we propose several deep learning baseline models for MMR; and finally, we explore the potential characteristics of memorable music pieces while providing interpretability for these deep learning-based methods.


%
\section{Related Work}
In addition to the cognition literature on music memorability~\cite{halpern2008effects,mullensiefen2012role}, there are several related yet distinct terms, such as Involuntary Musical Imagery (INMI) or "earworms"—fragments of music that involuntarily come to mind \cite{mccullough2015catching}. Studies have examined earworms through interviews, environmental and psychological conditions leading to INMI, and the impact of melodic features and song popularity on spontaneous musical imagination \cite{jakubowski2017dissecting}. Crucial differences between INMI and music memorability include: 1) INMI involves uncontrollable mental repetition, while memorability requires conscious recall; and 2) the stimuli in \cite{jakubowski2017dissecting} are highly familiar to participants, whereas our study selects audios unfamiliar to most annotators to mitigate the influence of individual listening histories on memorability. Another related concept is hook catchiness~\cite{burgoyne2013hooked, DBLP:conf/ismir/BalenBBMV15, van2016audio, korsmit2017if}, which refers to the most easily recalled fragment of a musical piece. However, our focus lies in predicting the memorability of different music pieces rather than assessing the impact of various segments within the same tune on catchiness prediction and recognition. Furthermore, we ensure our stimuli consist solely of pure instrumental music clips to prevent any textual information from lyrics influencing music memorability.

Moreover, while deep learning has achieved significant success in supervised MIR tasks, it often demands large-scale annotated data. However, collecting useful annotations for MIR tasks can be costly, as it typically requires expertise and domain knowledge \cite{mcfee2015software}. To tackle this challenge, various data augmentation and training strategies have been proposed~\cite{mcfee2015software,cubuk2020randaugment,wu2021multi,castellon2021calm}. For instance, McFee et al.\cite{mcfee2015software} apply transformations such as pitch shifting, time-stretching, and adding background noise to the original waveform. Cubuk et al.\cite{cubuk2020randaugment} mask both time and frequency content to expand the input space in automatic speech recognition (ASR) and MIR tasks. To enhance learning robustness with limited data, Wu et al.\cite{wu2021multi} extract general music representations using a multi-task pre-trained encoder, inspired by speech processing research \cite{Pascual2019, ravanelli2020multi}. Similarly, Castellon et al.\cite{castellon2021calm} employ transfer learning from existing music generation architectures. However, not all the aforementioned methods are simultaneously open-source, computationally inexpensive, and interpretable. Therefore, in this paper, we focus on applying signal processing approaches like masking, with further details provided in Section 4.


\begin{table*}
\centering
\small
\begin{tabular}{llllll}
    \hline
Task Type & \# of Audios & \# of Repetition (min) & \# of Repetition (max) & \# of Repetition (avg) & \# of Repetition (std)  \\ \hline \hline
Filler & 65 & - & - & - & - \\ \hline
Vigilance & 21 & 5 & 10 & 6.5 & 1.08  \\ \hline
Short-Term Target & 88 & 10 & 49 & 25.23 & 10.81  \\ \hline
Medium-Term Target & 41 & 61 & 131 & 110.33 & 16.32  \\ \hline
Long-Term Target & 20 & 155 & 276 & 222.05 & 36.64  \\ \hline
\end{tabular}
\caption{Details of different audio tasks in the music memory game.}
\label{tab:memory_game_overview}
\end{table*}


%

\section{Dataset construction for Music Memorability}
In this section, we discuss the details of our dataset collection process and how music memorability is measured.

\subsection{Audio Collection}
To construct a dataset with objective music memorability scores, we first ensure that the audio samples are unbiased. We randomly select music by querying music-related videos using the YouTube API with random query keys, avoiding any specific music genre preference. Additionally, we manually filter the music to confirm that the queried videos contain pure music content, excluding instrument tutorials or gadget unboxings. Next, we conduct a pilot study to verify that the selected audios are unfamiliar to most of the annotators in our target user group. Considering the annotators' nationalities might not be as varied as the music collection's, and language can be a memorable yet non-music-related element, we only use the intro part of each song. This approach helps eliminate other potential variables affecting music memorability. Also, the volume across all audio clips is normalized to minimize any memorable attributes unrelated to the music itself. Loudness normalization ensures the music is remembered based on its inherent qualities rather than its loudness.

We use only a segment of each audio for two reasons: i) to better eliminate confounding factors, such as vocal timbre, and ii) to shorten the period of annotations and prevent fatigue. We achieve this by truncating audios into structurally meaningful segments and applying proper time-stretching to alter the duration of an audio signal to a fixed length without distorting the audio. The segmentation process is supervised by an expert with a professional music education background. Note that time-stretching not only reduces modeling complexity but also prevents annotators from memorizing the audio based on its duration.

Ultimately, we collect 235 structurally meaningful 5-second audios with labeled music memorability scores. Our goal is to determine which types of music pieces are more likely to be memorized, rather than focusing on entire music clips, which are more complex and involve additional factors. This research can facilitate various applications, such as Netflix's iconic ``ta-dum'' sound. The collected data can be found in the supplementary materials. \figref{fig:source_dist} illustrates the distribution of the collected audios concerning their published geographical locations and total views on YouTube, with view counts ranging from 10K to 100M.

\begin{figure}[t]
 \centerline{
 \includegraphics[width=1\columnwidth]{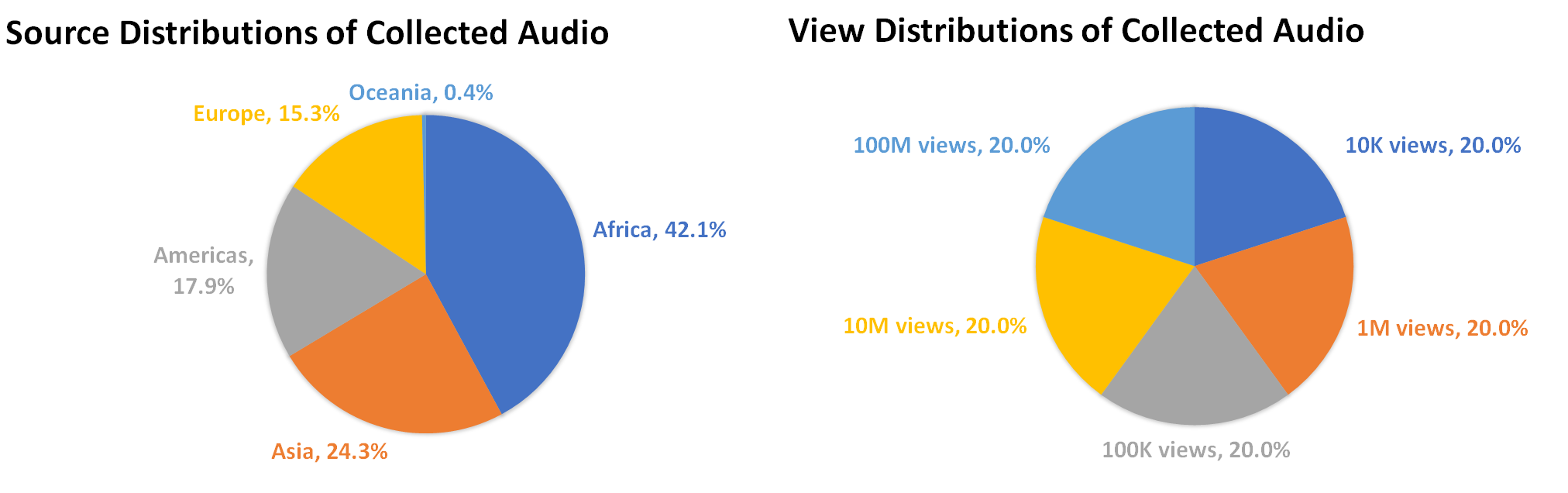}}
 \caption{Distributions of the audio published location and the distributions of the audio views in the final dataset.}
 \label{fig:source_dist}
 \vspace{-2mm}
\end{figure}

\subsection{The Music Memory Game}
To annotate the memorability of the collected musical data, we follow the setting of image memory game~\cite{isola2013makes} to design a novel music listening experiment. During the experiment, the recruited data annotators are asked to listen to 235 music pieces in total and answer whether the audio is repeated in the experiment or not. From a cognitive view, we define music memorability as long-term musical salience and the extent to which a musical piece continues to be remembered over time. In the music memory game, music memorability is measured as the tendency to correctly recognize a music piece when encountering it again in the experiment among all users. Specifically, let $x_j^{(i)}$ denote whether the $i$-th music piece can be recalled by the $j$-th data annotator, \textit{i.e.}, 1 if the annotator recognized the $i$-th music piece.
The memorability score of music $i$, denoted by $m^{(i)}$, is then calculated by: 
\begin{equation}
\label{eqn:equation1}
m^{(i)}=\frac{1}{n_{i}}\sum_{j=1}^{n_{i}}x_j^{(i)},\; x_j^{(i)}\in\{0,1\}
\end{equation}
where $n_{i}$ is the total number of data annotators for the $i$-th music.\\

To make the ground truth unbeknownst to all participants, music excerpts are split into three task categories: ``vigilance'', ``target'', and ``filler''. Targets and vigilance targets are both repeated in the experiment, while the former are collected to be the true labels and the latter is used to make sure participants are attentive when labeling data. Moreover, fillers are used to stuff the spacing between the first and second repetition of a target and therefore is only presented once. The overview of the music memory game experimenting procedure is shown in \figref{fig:exp_overview}. The target-vigilance-filler split details can be found in \tabref{tab:memory_game_overview}. Rigorous criteria are enforced to monitor the performances of data annotators and preserve the quality of memorability labels. Specifically, annotations from users who detect vigilance repetition with an accuracy lower than 60\% are automatically discarded.  Furthermore, to prevent gathering biased memorability, all annotators only engage in labeling once. We recruited a total of 218 users from campus, with 163 clearing the vigilance accuracy level, 17\% of passed annotators having professional music education backgrounds, and over 98\% being between the ages of 20 and 29. 

\begin{figure}
 \centerline{
 \includegraphics[width=\columnwidth]{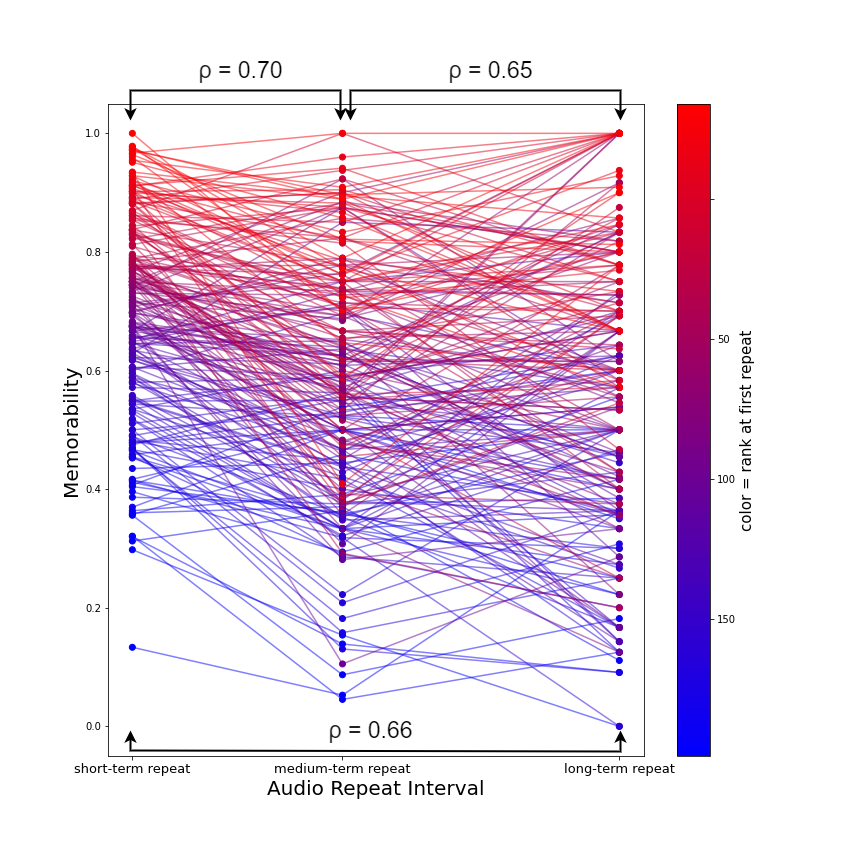}}
 \caption{Memorability scores at various stages. The color symbolizes the rank of short-term memorability, while the lines represent stage relationships. The plot also shows Spearman's rank correlations $\rho$ between memorabilities measured at each stage.}
 \label{fig:exp_stage_correlation}
\end{figure}

Differing from previous works on image memorability, our experiment is composed of three similar stages with breaks inserted in between. The reasons for using stages and breaks are two-fold. First, audios are sequential, therefore it is more exhausting to label the memorability score to audios as compared to static images. Second, it usually takes some time for the earworm phenomenon to happen when listening to music. Hence, we assume memorability should be invariant even after encountering breaks that probably would reset the memory. The results of relations between repeat interval and memorability score are shown in \figref{fig:exp_stage_correlation}, where the lines exhibiting memorability scores across short-term, medium-term, and long-term repeats. The results manifest that the memorability score is indeed independent of the sequential context. Therefore, it is easier to memorize truly memorable pieces of music even after long breaks. The fact that Spearman's rank correlation \cite{spearman1987proof} between short-term, medium-term, and long-term are all greater than 0.64 also proves that the rank of memorability score is preserved across variant repeat intervals.

\subsection{Labels and Consistency Analysis}
To assure that collected labels are universal across all data annotators, we evaluate the human consistency according to previous work \cite{isola2013makes} by randomly splitting all participants into 2 groups and examining how well the memorability scores measured in the first groups matched the ones measured in the second group by averaging Spearman's rank correlation \cite{spearman1987proof} between randomly separated two halves of the participants 25 times. The average Spearman's rank correlation coefficient $\rho$ is $0.83$, indicating the consistency of the collected data.

\begin{figure}
 \centerline{
 \includegraphics[width=\columnwidth]{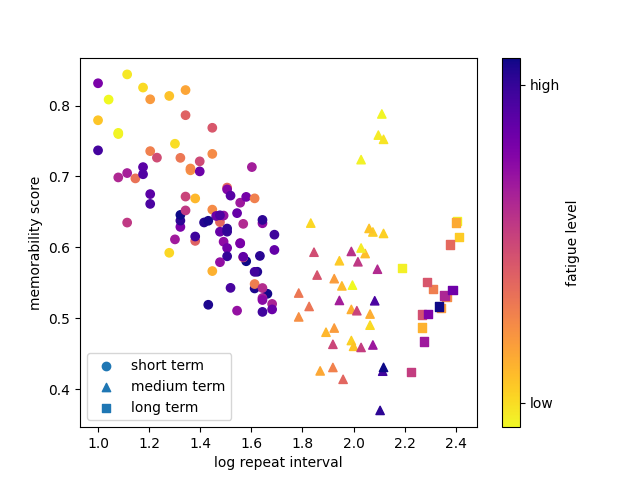}}
 \caption{Relations between memorability score and target repeat interval in log scale. The hue represents the level of fatigue.}
 \label{fig:score_to_repeat_length}
\end{figure}

\figref{fig:score_to_repeat_length} shows the scattering plot of music memorability and repeat interval. The graph demonstrates that music possesses a linear relation between memorability score and log-scaled repeat interval. Please note that the fatigue level is another factor in the plot that also contributes to the memorability score of audio. 
The fatigue level, defined as the amount of audios listened without a 3-minute break, is a direct result caused by staging experiment and participating in taking a break in the middle since listening to more music at one time without resting reduces participants' ability to identify repeated music pieces. The setting of inserting audio to random positions in the experiment procedure adds more context diversity to the process of memorizing music, thus making the labeled memorability scores more robust.

\section{Music Memorability Prediction}

\subsection{Learning with Handcrafted Features}
Although feature extractions for deep learning models can be data-driven without being handcrafted, leading to a better result given sufficient training data, handcrafted features provide interpretable information for more insights. Therefore, we propose handcrafted features that can more accurately depict the low-level acoustic features or high-level semantic features of musical clips as shown in \tabref{tab:tabular_features}. For the low-level acoustic features that can be directly derived from the audio signal of music segments, we leverage the harmony, rhythm and timbre since they are most easily recognizable fragments of a piece of music \cite{burgoyne2013hooked} and describe the fundamental elements of a tune. Moreover, zero crossings and zero crossing rate are also extracted since they give the impressions into the frequency content of a signal. On the other hand, high-level semantic features are more abstract descriptions. Since the previous works in Psychology \cite{samson2009emotional, vuilleumier2015musicemotion} mention the link between music emotion and memory, we introduce valence and arousal, which represent the mood of music pieces as features.

Another high-level feature is genre, which describes how likely a clip is belong to a certain type of music. Specifically, due to the unstable performance of existing algorithms for detecting sequences of chord labels, we employ chromagram (chroma) \cite{harte2005automatic} as a representation of harmony patterns. To extract timbre information, the Mel-Frequency Cepstral Coefficient (MFCCs) is widely utilized. Although MFCCs is representative for timbre, its components are difficult to grasp intuitively. As a result, we treat MFCCs as a raw feature and find an alternative solution by first separating source audios into four components using source separation software Spleeter~\cite{hennequin2020spleeter}, and calculating their respective amplitudes to represent the characteristics of different instruments and frequency ranges. For the rhythmic pattern, although Tempogram~\cite{cemgil2000tempo} captures the underlying rhythmic pattern of raw audios, it is unable to provide precise insights to concretely measure the audio's groove. Therefore, we instead utilize beat per minute (bpm) to represent general rhythm characteristics. We also use static valence and arousal values to describe perceived music moods, which are predicted by using Support Vector Regression (SVR) with a linear kernel trained on the PMEmo dataset~\cite{zhang2018pmemo}. For genre features, we use the predicted music tagging and instruments from the downstream task of PANN \cite{kong2020panns}. Finally, SVR and Multilayer Perceptron (MLP) are employed as predictors to link audio features to memorability scores.

\begin{table}[t]
\centering
\small
\begin{tabular}{p{0.08\textwidth}p{0.11\textwidth}p{0.22\textwidth}}
\\
    \hline
Level & Category & Feature Implementation  \\
    \hline \hline
\multirow{4}*{Low-level} & Harmony  & mean, std of 12 pitch class \\
~ & Rhythm   & beat per minute (bpm)  \\
~ & Timbre   & mean, std of 4-tracks (Vocals, Bass, Drums, Others)   \\
~ & Zero Crossing & \# of zero crossings \& avg, median of zero crossings rate\\
    \hline
\multirow{2}*{High-level} & Mood & valence, arousal \\
~ & Genre &  Music, Musical Instrument\\
\hline
\end{tabular}
\caption{Explainable handcraft features.}
\label{tab:tabular_features}
\end{table}

\subsection{End-to-End Deep Learning}
Deep learning~\cite{lecun2015deep} is featured by its ability to directly learn meaningful information from raw data, as opposed to using hand-crafted features. As a result, we also test end-to-end models to find if their feature-learning process improves performance. Our model uses spectrograms in Mel-scale as inputs, similar to previous end-to-end MIR tasks. Moreover, transfer learning~\cite{pan2009survey}, which applies previously learned knowledge to new data, has been found to significantly increase learning performance by skipping costly data-labeling procedures. Here, we use the self-supervised pre-trained Audio Spectrogram Transformer (SSAST) \cite{gong2021ssast} since SSAST has been proved to achieve state-of-the-art results on numerous audio tasks, including audio event classification, keyword spotting, mood recognition, and speaker identification, after being trained on a vast amount of unlabeled data.

\section{Experiment Results}
\noindent\textbf{Evaluation Metrics.} Spearman's rank correlation and mean squared error (MSE) loss are used as the metrics to evaluate the performance of music memorability prediction. The former indicates the ability to rank the relative memorability of different audios, while the latter indicates the absolute error of the predicted results.

\noindent\textbf{Different Baselines.} Here, we leverage Chroma and MFCCs along with their respective derivatives as two hand-crafted feature representations and fit the ground truth by Multilayer Perceptron (MLP) as two simple baselines. Moreover, we also use the convnet model as a baseline since it is the most referenced and available work in general music representation. The convnet model \cite{choi2017transfer} utilizes CNNs for music tagging in the pre-training stage, and the extracted features serve as the representation for downstream tasks. Finally, Self-Supervised Audio Spectrogram Transformer (SSAST)~\cite{gong2021ssast} is also used as the baseline, which is a Transformer-based model with more parameters as compared to CNNs. SSAST pretrains the model with joint discriminative and generative masked spectrogram patch modeling. 

\noindent\textbf{Implementation Details.} All the feature classifiers are pretrained without finetuning on the self-collected dataset. To handle the instability stemming from the limited labeled data, we normalize labels by subtracting the mean value, \textit{i.e.}, predicting a relative value instead of an absolute value. For MLP and SSAST models, the learning rates are respectively set to 5e-5 and 5e-6 with the Adam optimizer~\cite{DBLP:journals/corr/KingmaB14}. We also conduct additional feature selection on the handcrafted features to improve the convergence of the MLP/SVR model (only select 25 features) due the small number of data samples. In addition, techniques including frequency masking, band stop filtering, and reverberation \cite{ravanelli2020multi} are used to augment data, together with the pitch shifting augmentation. The results are reported by the average of the 10-fold outputs.

\begin{table}
\centering
\small
\begin{tabular}{llll}
    \hline
Method                & Corr. & MSE  & MSE STD   \\
    \hline \hline
chroma + MLP          & 0.1740  & 0.0326 & -  \\
    \hline
MFCCs + MLP            & 0.1179  & 0.0353 & - \\
    \hline
convnet features~\cite{choi2017transfer} + MLP & 0.1889       & 0.0314 & - \\
    \hline
EHC features + SVR   & \textbf{0.2988}   & 0.0339  &  0.0128 \\
    \hline
EHC features + SVR + PS & 0.2084  & 0.0391 &  0.0129 \\
    \hline
EHC features + MLP   & 0.2656   & 0.0263  &  0.0058 \\
    \hline
EHC features + MLP + PS & 0.2388  & 0.0275  &  0.0059 \\
    \hline
mel-spectrograms + SSAST   & 0.0124  & 0.0298  &   0.0061 \\
    \hline
mel-spectrograms + SSAST + PS  & 0.2658  & 0.0265  & 0.0074  \\
    \hline
\end{tabular}
\caption{Spearman's rank correlation and MSE loss between predicted and ground truth music memorability score using different models. Note that EHC features stand for explainable handcrafted features, PS stands for pitch shift data augmentation, and Corr. represents Spearman's rank correlation.}
 \label{tab:exp_results}
\end{table}


\subsection{Prediction Results.}
\tabref{tab:exp_results} compares the results of different prediction models, where SVR and MLP take explainable handcrafted features as inputs, and SSAST takes mel-spectrograms as inputs. The results indicate that chroma and MFCCs produce the worst results due to the ineffective feature extraction. For convnet features, the performance is better than chroma and MFCCs due to the pretraining. However, the amount of training data is too small to finetune the model on the music memorability regression task. SSAST outperforms other baselines since it incorporates the prior knowledge of spectrograms pre-trained by using advanced methods. Finally, Explainable Handcrafted Features (EHC) method produces the best correlation results by combining both low- and high-level features that help improve music memorability. These quantitative findings manifest that data-driven MIR tasks are notably reliant on huge data quantities to be resilient and general.

\subsection{Ablation Study}
Table \ref{tab:featureSelect_results} shows an ablation study on feature selection for handcrafted features, indicating that selecting top-25 features leads to the best overall correlation results. Moreover, Table \ref{tab:exp_results} also shows an ablation study on extra pitch shifting for data augmentations. Small pitch shifts (less than 5 semitones) make the altered audio seem natural to the human ear according to \cite{thickstun2018invariances}. Therefore, we add semitone shifts of -5 to 5 to our data. The results manifest that pitch shifting is effective for the models that take sequence information into account because applying mean pooling across time on harmony features in non-sequential models like SVR and MLP just forces the model to forecast the same value using multiple static chroma information. This may confuse the model on harmony characteristics. On the other hand, models with sequential information, such as SSAST, learn pitch-invariance after pitch shifting. The performance of SSAST notably decreases without pitch shift data augmentation, possibly due to its data-hungry nature as a Transformer-based model, \textit{i.e.} requiring more data for optimal parameter tuning.

\begin{table}
\centering
\small
\begin{tabular}{llll}
    \hline
Model & Top-k feature selection & Corr. & MSE  \\ \hline \hline
MLP & k = 40 (no feature selection) &  0.2160  & 0.0272  \\ \hline
MLP & k = 35  & 0.2324 & 0.0270  \\ \hline
MLP & k = 25  & \textbf{0.2656} & \textbf{0.0263}  \\ \hline
MLP & k = 20  & 0.2018 & 0.0271  
\\ \hline \hline
SVR & k = 40 (no feature selection) &  0.2168 & \textbf{0.0324}  \\ \hline
SVR & k = 35 &  0.2291  & 0.0340 \\ \hline
SVR & k = 25 &  \textbf{0.2988}  & 0.0339  \\ \hline
SVR & k = 20 &  0.2630  & 0.0354  \\
\hline
\end{tabular}
\caption{Spearman's rank correlation and MSE loss for MLP/SVR models with different top-k feature selection.}
 \label{tab:featureSelect_results}
\end{table}


\subsection{Interpretability} 
\begin{figure}
 \centerline{
 \includegraphics[width=\columnwidth]{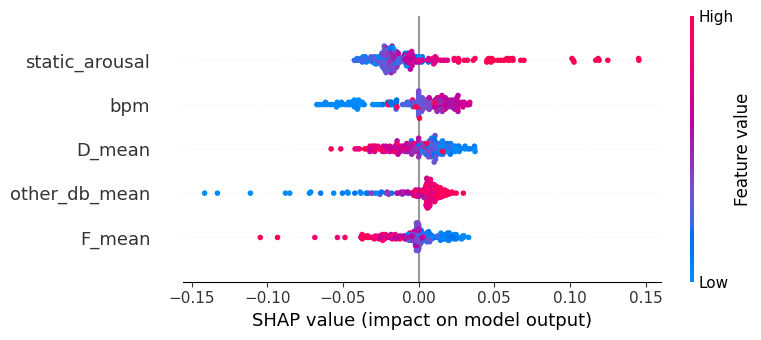}}
 \caption{SHAP summary of the SVR model with RBF kernel \cite{rbf1705021}. The most important features are listed in decreasing order and the fact that feature value rises after the SHAP value shows a positive relationship between the two.}
 \label{fig:shap_summary}
\end{figure}

We attempt to gain insight into the intrinsic memory utilizing XAI methodologies. One post-hoc strategy for expressing black box models in a human-interpretable manner is SHAP~\cite{NIPS2017_7062}. Specifically, SHAP explanations are obtained by perturbing a specific instance in the data and observing the impact of these perturbations on the black-box model's output. As such, SHAP allows us to explore the factors that the model considers when determining memorability. \figref{fig:shap_summary} visualizes the directionality impact of the top-5 features in SHAP, where the x-axis stands for SHAP value and each point is a SHAP value of a sample for a feature. Red color and blue colors respectively indicate higher and lower values of a feature. As such, we can observe the feature directionality impact based on the distribution. For example, the first row shows that a higher arousal value leads to high memorability scores, while a lower arousal value can lead to both high and low memorability scores. The important factors for the predictor among the EHC features include arousal, bpm, harmony (the feature "D mean") and the timbre features extracted from the source other than vocals, drums, and bass (the feature "other db mean"). According to Psychology research \cite{samson2009emotional}, normal individuals without brain damage find it easier to recognize musical excerpts with high arousal. The melodies are the main constituent elements of the source ''others'' after applying 4-stem Spleeter separation. This finding supports our understanding that we often focus on the main melody in music, and thus the chorus or hook of the song with outstanding melody usually represents the entire song.

\section{Conclusion and Future Work}
In this work, we explore the novel task of music memorability regression (MMR) using a data-driven approach. The consistency of our newly proposed YouTube Music Memorability (YTMM) dataset supports our hypothesis that music memorability indeed exists and can be predicted. Furthermore, we investigate the use of feature engineering and self-supervised learning for predicting music memorability, highlighting the importance of prior knowledge and other training approaches, such as label normalization, for improving results with limited data. We make the dataset available online to encourage further research and development in the field of MMR. In the future, we plan to: 1) scale the dataset to better represent the memorability of full music structures, 2) investigate the potential of transfer learning trained on music-oriented datasets to enhance our current baselines, and 3) study the personalization issue since music memorability can be strongly related to the past musical experience of individuals.


\section{Acknowledgement}
\label{sec:ack} 
This work was supported in part by the National Science and Technology Council of Taiwan under Grant NSTC-109-2221-E-009-114-MY3.

\bibliography{ISMIRtemplate}

%
%
%
%
%

\end{document}